\documentstyle[12pt]{article}
\begin{document}
\title{ONE LOOP CALCULATIONS ON THE WESS-ZUMINO-WITTEN ANOMALOUS
FUNCTIONAL
AT FINITE TEMPERATURE}
\vskip 1.0cm
\author{{\large\bf R.F. Alvarez-Estrada}, {\large\bf  A. Dobado} and
{\large\bf A. G\'omez Nicola}
\\Departamento de F\'{\i}sica
Te\'orica, Universidad Complutense
\\28040, Madrid, Spain}
\maketitle
\begin{abstract}
We analyze the finite temperature (T) extension of the Wess-Zumino
-Witten
functional, discussed in a previous work, to one loop in chiral
perturbation theory. As a phenomenological application, we calculate
finite temperature corrections to the amplitude of $\pi^0$ decay
 into two photons. This calculation is performed in
three limits : i) $T/M_{\pi}<<1$ , ii) the chiral limit at finite T
and iii) $T/M_{\pi}>>1$ ($M_{\pi}$ being the pion
mass). The $T$-corrections
tend to vanish in the chiral limit, where only  the kaon
contribution
remains (although it is exponentially suppressed).
\end{abstract}
\vskip 2.0cm
FT/UCM/15-93
\baselineskip 0.83 true cm
\textheight 20 true cm

\newpage
The extension of the Wess-Zumino-Witten (WZW) anomalous
lagrangian \cite{wz-wit-agg}
for the finite temperature ($T\neq 0$)
case has been analyzed recently \cite{adg}. Instead of the
usual $S^4$
compactification  of Euclidean space-time at $T=0$
\cite{wz-wit-agg},
 a $S^1\times S^3$ one was employed for $T\neq 0$.
Our task  is now to evaluate one-loop effects
using the $S^1\times S^3$ lagrangian,
in the trivial (pionic) sector with baryon number
$N_B=0$. In particular, we are interested in
the finite temperature corrections to the amplitude (or the
correlation function) corresponding to the decay $\pi^0
\rightarrow\gamma\gamma$. For the trivial sector, the finite T
 WZW action gauged
with the electromagnetic field $A_{\mu}$,
can be obtained directly from the one discussed in \cite{adg}.
It reads:
\begin{eqnarray}
\Gamma_{WZW}[U,A] &=& \Gamma_0[U] -e\int_{S^1\times S^3} A_{\mu}J^{\mu} +
\frac{e^2}{24\pi^2}N_c \int_{S^1\times S^3} \epsilon_{\mu\nu\alpha\rho}
\partial_{\mu}A_{\nu}A_{\alpha}Tr [Q^2(\partial_{\rho}U)U^{-1}+ \nonumber
\\
&+& Q^2 U^{-1}
(\partial_{\rho}U)+ \frac{1}{2}Q\partial_{\rho}UQU^{-1}-
\frac{1}{2}QUQ\partial_{\rho}U^{-1}] \nonumber \\
J_{\mu}&=& \frac{N_c}{48\pi^2}\epsilon^{\mu\nu\alpha\rho}Tr
[Q(\partial_{\nu}UU
^{-1})(\partial_{\alpha}UU^{-1})(\partial_{\rho}UU^{-1})+ \nonumber \\
&+&Q(U^{-1}\partial_{\nu}
U)(U^{-1}\partial_{\alpha}U)(U^{-1}\partial_{\rho}U)] \nonumber \\
\Gamma_0 [U]&=&\frac{N_c}{240\pi^2}\int_{0}^{1} dt\int_{S^1\times
S^3} \epsilon^{ijklm}U_t^{-1}\partial_i U_t U_t^{-1}\partial_j U_t
 U_t^{-1}\partial_k U_t U_t^{-1}\partial_l U_t U_t^{-1}\partial_m U_t
\label{eq.1}
\end{eqnarray}
where $i,j,k,l,m=1,...5$, $t\in[0,1]$, and $U_t$ being a mapping from
$[0,1]\times S^1\times S^3$ into $SU(3)$ with $U_0=1$,
$U_1=U(\vec{x},\tau)$.
In turn,
$U\equiv U(\vec{x},\tau)$ is a mapping  from $S^1\times
S^3$ to $SU(3)$, which can be parametrized as $U(\vec{x},\tau)=
\exp(i\pi^a(\vec{x},\tau)T_a/F_{\pi})$ with $a=1 ... 8$
and  $\pi^a(\vec{x},\tau)$
being the Goldstone boson fields ($\pi$, $K$, $\eta$). $T_a$
are the $SU(3)$ generators
and $F_{\pi}\simeq 93$ MeV is the pion decay constant. Notice
that $\Gamma_0[U]$
is the direct restriction of eq.(14) in \cite{adg} to the
trivial ($j=0$)
sector. The trace in (\ref{eq.1})
is over $SU(3)$, $\beta=1/T$
is the radius of $S^1$, $N_c$ is the number of colours,
and $Q$ is the quark charge
matrix:
\begin{equation}
Q=\left( \begin{array}{ccc}
\frac{2}{3}&0&0 \\
0&-\frac{1}{3}&0 \\
0&0&-\frac{1}{3} \end{array} \right )
\label{eq.2}
\end{equation}

At low energies, the  total action of the system is obtained by
adding to the WZW term
(\ref{eq.1}) the non-linear sigma model piece in the Euclidean
$S^1\times S^3$ space:
\begin{eqnarray}
\Gamma&=&\frac{1}{4}F_{\pi}^2\int_{S^1\times S^3}
Tr(D_{\mu}U^{-1}D^{\mu}U-M^2(U+U^{-1}))+\Gamma_{WZW}
\nonumber \\
D_{\mu}&=&\partial_{\mu}+ie[Q,U]A_{\mu}
\label{eq.3}
\end{eqnarray}
where $M$ is the quark mass matrix. Using standard techniques,
one generates
the  Chiral Perturbation Theory ($\chi $ PT)
 expansion. Higher energy or temperature contributions
can be included by adding higher derivative terms to the above action.
 Since we are interested only in the low temperature domain
 those terms will not be considered here. Note also that working
 at finite $T$
the $\chi PT$ expansion is done in powers of the parameter
$T^2/8F_{\pi}^2$
\cite{ge-le}.

Now we concentrate on the $\pi^0
\rightarrow\gamma\gamma$ process.
  In general, the  $\pi^0$ decay amplitude at any $T$ can be expressed
 as:
 \begin{equation}
 A_{\mu\nu}=\frac{iN_c e^2}{12\pi^2 F_{\pi}}C(T)
 \epsilon_{\mu\nu\alpha\rho}
 k_2^{\alpha}k_3^{\rho}
 \label{eq.4}
 \end{equation}
where
$k_2$ and $k_3$ are  the fourmomenta of the outgoing photons and
$C(T)$ is
 some suitable
 function. At $T=0$ the lowest order approximation can be obtained
 from the WZW action at the
tree level and corresponds to $C=1$. Next order contributions come
from the
one-loop diagrams a) and b) in Fig.1
 (plus some renormalization graphs to be discussed
later). Those diagrams  have been studied at $T=0$
  in \cite{do-wy} \cite{do-ho-li}. Let us now
 consider those graphs together with
the usual renormalization diagrams for $F_{\pi}$ (graph c) in
Fig.1) and for the pion mass and wave function ( graph d) in Fig.1).
 Then, the final result for the one-loop $\pi^0\rightarrow\gamma\gamma$
 amplitude  is the same as the tree level one, provided that $F_{\pi}$ be
 replaced by its
 physical value, for on-shell photons. In other words it is given by
(\ref{eq.4}) also with $C(0)=1$. Let us sketch its justification.
The contributions of the graphs
a), c) and d) in Fig.1 are proportional to $G_{\pi}(0)$ and
$G_{K}(0)$, $G_{\pi}(x)$ and $G_{K}(x)$ being the pion and
kaon propagators ($x \in S^4$).
At $T=0$ it can be shown that (in dimensional
regularization) the only surviving piece of the graph b)
that contributes to
the decay amplitude is a sum of terms  proportional to $G_{\pi}(0)$ and
$G_{K}(0)$. The
corresponding coefficients of every contribution cancel with one other,
for pions and kaons separately. Notice that $G_{\pi}(0)$ and
$G_{K}(0)$ are divergent quantities, so that the above cancellations
imply that there are no one-loop
divergent contributions to the anomaly that could renormalize the $N_c$
coefficient   of the WZW action \cite{do-wy}.

Now we turn to analyze the $T\neq 0$ case. Physically, it would
correspond to the effect of a medium at constant temperature $T$, in
which the $\pi^0$ could live. Such a situation could take place in
a neutron star, or in a cosmological hadronic medium below the  chiral
phase transition point \cite{ge-le} \cite{gatto} \cite{sch}.
In agreement with
previous studies of the $\pi^0\rightarrow\gamma\gamma$ decay at finite
temperature \cite{colo-ag}, one expects to find non-vanishing
$T$-corrections. In addition  such corrections should be
ultraviolet finite and  no renormalization of the anomaly
at finite $T$ should be needed.

The imaginary time formalism of Finite Temperature  Field Theory
(FTFT) \cite{la-va} can be implemented by using
 the $S^1\times S^3$ compactification  of the WZW action discussed
in \cite{adg}. As usual the     energies
are discretized and
 energy integrals must be replaced by sums over Matsubara frequencies.
 Let us concentrate on the contribution of graph b) in Fig.1. It splits
 into
a pionic and a kaonic part, depending on the particle inside the loop.
{}From the
lagrangian in (\ref{eq.1}) and (\ref{eq.3}), the pionic part  reads in
Euclidean  momentum space :

\begin{eqnarray}
A_{\mu\nu}^{(b)}(T)&=&\frac{iN_c e^2}{3\pi^2 F_{\pi}}\frac{1}{F_{\pi}^2}
\epsilon_{\nu\alpha\rho\gamma}k^{2\gamma}k^{3\alpha}I_{\mu}^{\rho}(T)
\nonumber
  \\
I_{\mu}^{\rho}(T)&=&\frac{1}{\beta}\sum_{n=-\infty}^{+\infty}\int
\frac{d^{d-1}\vec{q}}{(2\pi)^{d-1}}
\frac{q_{\mu}q^{\rho}}{(\omega_n^2 + \vec{q}^2+M_{\pi}^2)
((\omega_n-k_{20})^2+
(\vec{q}-\vec{k_2})^2+M_{\pi}^2)}\nonumber\\
&&
\label{eq.5}
\end{eqnarray}

where $\omega_n=2\pi n/\beta$, $d=4-\epsilon$ (see \cite{ge-le})
and $M_{\pi}$
is the pion mass. The kaon term has the same structure as (\ref{eq.5})
with
$M_{\pi}$ replaced by $M_{K}$. Now, by using the well known
formulae:
\begin{equation}
\frac{1}{AB}=\int_0^1 dx \frac{1}{[Ax+B(1-x)]^2}
\label{eq.6}
\end{equation}
and
\begin{equation}
\frac{i}{\beta}\sum_{n=-\infty}^{+\infty} f ( \frac{2n\pi i}
{\beta}) = \frac{1}{2\pi}\sum_{s=\pm1} \int_{-i\infty+s\epsilon}
^{i\infty+s\epsilon}\frac{d\omega f(\omega)}{\exp[s\beta\omega]-1}
 +\frac{1}{2\pi} \int_{-i\infty}
^{i\infty} d\omega f(\omega)\quad ;\epsilon\rightarrow 0^+
\label{eq.7}
\end{equation}
in (\ref{eq.5}) we can separate the
$T=0$ part and arrive to
an integral over the $\omega$ complex plane for the $T\neq0$ piece. This
integral can be performed by
using the residue theorem inside the contours $C_1$ and $C_2$ showed in
Fig.2. In fact, the resulting $f(\omega)$ has always two
poles at the points $\omega=ik_0\pm B(\vec{q};x)$ with
$B(\vec{q};x)=\sqrt{A(\vec{q};x)-k_0^2}$, $A(\vec{q};x)=\vec{q}^2-
2\vec{q}\cdot\vec{k}+M^2 (x)$, $M^2 (x)=M_{\pi}^2+(1-x)(k_{20}^2+
\vec{k_{2}}^2)$ and $k_{\mu}=(1-x)k_{2\mu}$. One of those poles lies
always inside
 $C_1$ and the other one inside $C_2$.

In the following, in order to make the discussion simpler
 we will consider three different limiting cases, namely
 the low temperature limit $T/M_{\pi}<<1$, the chiral limit $M_{\pi}=0$
 and finally
the  limit $M_{\pi}/T<<1$. Note that this last limit
 can be understood as the one caracterizing the  leading corrections
 to  the previously mentioned chiral limit
$M_{\pi}=0$.

We shall consider first the low temperature limit $\frac{T}{M_{\pi}}<<1$.
In this regime, the $T\neq 0$ piece in  $I_{\mu}^{\rho}(T)$  can
be written
, as $d\rightarrow 4$, as a linear combination of   the one-dimensional
integrals:

\begin{equation}
I_k(n;T,x)=\frac{2}{(4\pi)^{\frac{k-1}{2}}\Gamma(\frac{k-1}{2})}
\int_0^{\infty}dy \frac{y^{k-2}}{(y^2+M^2(x))^{n/2}}\exp\left(\frac{
-\sqrt{y^2+
M^2 (x)}}{T}\right)
\label{eq.8}
\end{equation}

The integrals in (\ref{eq.8}) are always finite for $M_{\pi}\neq 0$.
Then we see that at low temperatures
there are no divergent temperature corrections for $d\rightarrow 4$,
in agreement
with the idea that renormalization at $T=0$ ensures that no extra
divergent
terms appear at $T\neq 0$. In particular, that  is also true for
off-shell
photons with $k_{20}^2+\vec{k_{2}}^2\neq 0$
(after analytic continuation of $k_{20}$
to continuous values), as studied in \cite{do-wy} for $T=0$.

In order to study the finite $T$ corrections to the $\pi^0$
decay amplitude, we consider: i) the usual analytic continuation
from discretized
to continuous frequencies of the external photons, ii) on-shell photons
($M^2 (x)=M_{\pi}^2$ and  $I_k(n;T,x)$ being, then, independent of $x$)
and
iii) the $\pi^0$ rest frame, in which $k_{2j}=-k_{3j}$, $k_{20}=k_{30}=
iM_{\pi}/2$. On the other hand, the contributions of diagrams a),c) and
d) of Fig.1 are now proportional to $G^T_{\pi}(0)$. In turn
$G^T_{\pi}(x)$ is
the thermal pionic propagator (now $x\in S^1\times S^3$).
One has \cite{ge-le} :

\begin{eqnarray}
G^T_{\pi}(0)&=&G_{\pi}(0)+g_1 (M_{\pi},T) \nonumber \\
g_1 (M_{\pi},T)&=&2\sum_{n=1}^{\infty}\int_0^{\infty}
\frac{d\lambda}{(4\pi\lambda)^{d/2}}
\exp (-\lambda M^2_{\pi}-\frac{n^2}{4\lambda T^2})
\label{eq.9}
\end{eqnarray}

In the $\pi^0$ rest frame, the only surviving contribution of
$I_{\mu}^{\rho}$
in (\ref{eq.5}) is
the $I_{i}^{j}$ part ($i,j=1 ... d-1$). This part is splitted
into two pieces,
one of them  proportional to $\delta_{i}^{j}$ and the other one to
$k_{2i}k_{2}^{j}$.
The later vanishes in $A_{\mu\nu}^{(b)}(T)$ in (\ref{eq.5}).
The $\delta_{i}^{j}$ piece is written in terms of the $I_k$
integrals in
(\ref{eq.8}).
 The final result for the $\pi^0$ decay amplitude at low temperature
 (having
accomplished the steps i),ii) and iii) above and letting $d\rightarrow 4$)
is
given in (\ref{eq.4}), with:

\begin{eqnarray}
C^{RF}(T)&=& 1+\frac{2}{F_{\pi}^2}(2f(M_{\pi},T)-g_1(M_{\pi},T))+
{\normalsize (kaon \quad terms)} \nonumber \\
f(M_{\pi},T)&=&\frac{2\pi}{M_{\pi}}\sinh(\frac{M_{\pi}}{2T})
(I_6(2;T)+TI_6(3;T))
\label{eq.10}
\end{eqnarray}

$I_6(2;T)$ and $I_6(3;T)$ are given by the right-hand-side of
(\ref{eq.8})
with $M^2(x)=M^2_{\pi}$. The superindex RF refers to the pion
rest frame.
{}From (\ref{eq.8}-\ref{eq.10}) we obtain
the following
low temperature expansion for $C^{RF}(T)$  :

\begin{eqnarray}
C^{RF}(T)&=&1+2\frac{T^2}{F_{\pi}^2}\frac{1}
{(2\pi)^{3/2}}\tau^{-1/2}e^{-1/\tau}
(2\tau\sinh\frac{1}{2\tau}-1)\{1+\frac{3}{8}
\tau-\frac{15}{128}\tau^2 +\nonumber \\
&+&O(\tau^3)\} + O(e^{-2/\tau})+{\normalsize (kaon \quad terms)}
\label{eq.11}
\end{eqnarray}
where $\tau=T/M_{\pi}$. As
the exponentials in the above equation clearly suggest, the contribution
of
other families (the kaon terms), is
exponentially small
at low temperatures. The same occurs for other thermodynamical quantities
\cite{ge-le}.

Let us now analyze what happens in the chiral limit $M_{\pi}\rightarrow 0$
at
finite T.
First it is important to notice that one should take the infinite volume
 limit, $R\rightarrow
\infty$, $R$ being the radius of the $S^3$ sphere, before the
$M_{\pi}\rightarrow 0$ limit is taken. This is so
because the chiral limit would lead to inconsistencies if $R$
would have remained finite, since there are not massless Goldstone bosons
at finite volume
 (see \cite{leut}).
 From current algebra
theorems at $T=0$, one expects the $\pi^0$ decay amplitude to be
entirely dominated
by the anomaly in the chiral limit, and then, the low energy corrections
are expected to vanish. As we have derived the finite T corrections
displayed
in (\ref{eq.10})-(\ref{eq.11})  from a low energy action,
they are expected to be vanishingly small in the chiral limit
by the previous argument,
  as the anomaly does not depend on temperature.
Indeed, this is the case, which constitutes an interesting check of
consistency
of our calculations. In fact, it is possible to   calculate this limit
 for $I_{\mu}^{\rho}(T)$ in (\ref{eq.5})
by using again (\ref{eq.6}) and (\ref{eq.7}).
 Then we obtain for $d\rightarrow 4$ and
on-shell photons:

\begin{eqnarray}
I_{\mu}^{\rho}(T,M_{\pi}=0)&=&I_{\mu}^{\rho}(0,M_{\pi}=0)-
g_{\mu}^{\rho}\frac{T^2}{12}
+k_{2\mu}k_{2}^{\rho}{\normalsize - terms}
\label{eq.12}
\end{eqnarray}

The contribution of the $k_{2\mu}k_{2}^{\rho}$ terms in (\ref{eq.12}) to
$A_{\mu\nu}^{(b)}(T)$ in
(\ref{eq.5}) vanishes. The $g_{\mu}^{\rho}$ piece
cancels with the contribution coming from $g_1(M_{\pi},T)$, which is
the massless
 Bose gas self-energy
(see \cite{ge-le}). So we find that the only  $T$ contribution
 in the chiral limit is the one coming from the kaonic part. The
 latter is
exponentially diminished as long as the  temperature lies below the
kaon mass:

\begin{equation}
C(T,M_{\pi}=0)=1+O(\frac{T^2}{F_{\pi}^2}\frac{T^3}{M_K^3}
e^{-\frac{M_{K}}{T}})
\label{eq.13}
\end{equation}

Let us now come to the $T/M_{\pi}>>1$  regime.
The chiral limit contribution previously analyzed can also be viewed as
the leading one in this limit.
In the $M_{\pi}\neq 0$ case, it is possible to find higher order
contributions
in this limit, taking into account that the parameter $T^2/8F_{\pi}^2$
must
remain
small (in order to keep the  one-loop approximation in $\chi$PT valid).
 For this calculation, we leave the sum over frequencies
 (over n) as it appears in (\ref{eq.5}), without going to the complex
 plane.
Only (\ref{eq.6}) is then used in (\ref{eq.5}). The integral
over the $d-1$ momentum space can then be reduced to a one-dimensional
integral
by using standard dimensional regularization formulas. By expanding in
terms
of $\sigma=M_{\pi}/(2\pi T)$, it is found that the $n$-sum in every term of
  the expansion is of the form $\sum_{n=1}^{\infty} 1/n^{p+\epsilon}$
 $=\zeta(p+\epsilon)$ with $\zeta$ the Riemann's zeta function and
$\epsilon=4-d$. Notice that in this
treatment, the $T=0$ part has not been separated from the $T\neq 0$ one up
to this stage. The $p=1$ contribution in $\zeta$ carries a $1/\epsilon$
infinite
term which is nothing but the $T=0$ one.
Once this $T=0$ part is separated, we again find  that the remaining
$T\neq 0$
part is finite. In particular, as the first order term in the expansion
we find the $M_{\pi}=0$ term discussed before. Collecting the terms
coming from
 $g_1 (M_{\pi},T)$ that have already been calculated for $\sigma<<1$ in
\cite{ge-le}, we  have
for $C^{RF}(T)$ in this regime:

\begin{eqnarray}
C^{RF}(T)&=&1+\frac{T^2}{F_{\pi}^2}(c_1 \sigma+c_2 \sigma^2 + O(\sigma^4)+
O(e^{-\frac{1}{T}(M_{K}-\frac{M_{\pi}}{2})}))\nonumber\\
c_1 &=& 1-\frac{\sqrt{3}}{4}-\frac{\pi}{6}\qquad   c_2 = \frac{-1}{12}
\label{eq.14}
\end{eqnarray}

Notice that the $O(\sigma^2\log\sigma)$ terms  appearing in the expansion
of $g_1 (M_{\pi},T)$ \cite{ge-le} cancel exactly with the ones coming from
the graph b).

We now compare  the different behaviours
 of $C^{RF}(T)$ for low and high temperatures
 by looking at eqs. (\ref{eq.11}) and (\ref{eq.14}). Let us define $s(\tau)$
through $C^{RF}(T)=1+\frac{T^2}{F_{\pi}^2}s(\tau)$ ($s(\tau)$ being
dimensionless).
In Fig.3 we have plotted $s(\tau)$ versus $T$ (for $M_{\pi}$= 135 MeV), up to
temperatures of 200 MeV (which is an interval where our approximations
are expected to be valid).
 The kaon contribution has been plotted in the $T/M_{K}<<1$ limit.
In order to check the consistency of the several approximations considered,
we have also evaluated numerically the $T\neq 0$ part in
$I_{\mu}^{\rho}(T)$ in (\ref{eq.5}), through the exact expressions of the
residues, valid for any $T$. The results for some values of $T$ are also
showed in Fig.3. We see that the analytical results are in good
 agreement with the numerical computations in the two limits considered.

It is interesting to observe that for low
temperatures $s(\tau)$ is an increasing function of $\tau$,
while, in the large $T$ regime, it decreases with $\tau$.
 Whether or not
 this change of behaviour of the decay amplitude
 could be due to the presence of the chiral phase transition,
 near  the
temperature
range in which $s(\tau)$ decreases,
stands as an open question ( see other studies of
the chiral phase transition for
$M_{\pi}\neq 0$ like \cite{ge-le}\cite{gatto}).

Summarizing, the finite-T WZW functional \cite{adg},  gauged
with the electromagnetic field in the trivial sector, has been explored
up to one-loop in $\chi$PT.
 The resulting finite-T corrections to the $\pi^{0}\rightarrow\gamma\gamma$
decay amplitude have been analyzed in three regimes : i)$T<<M_{\pi}$,
ii)$M_{\pi}=0$ at finite T and iii)$T>>M_{\pi}$. As a check of consistency,
 we have found that the $T$-corrections to the decay amplitude tend to
 vanish in the chiral limit,
where only a exponentially supressed kaon contribution remains.

The financial support of CICYT (Proyecto AEN93-0776) is acknowledged.

\pagebreak
\makebox[5in]{\Large \bf FIGURE CAPTIONS}
\linebreak
\begin{description}
\item{\em Fig.1.}Diagrams contributing to the $\pi^0$ decay amplitude.
Graph c) gives rise to $F_{\pi}$ renormalization
 and d) is the pion mass and wave function renormalization graph . The
dashed line in graph c) represents the axial current
(see \cite{do-wy}\cite{do-ho-li}).
\item{\em Fig.2.}Contours $C_1$ and $C_2$ in the $\omega$ complex plane.
\item{\em Fig.3} $s(\tau)$ as a function of $T$ for $M_{\pi}=135$ MeV. The
two regimes (high T and low T) analyzed in the text are plotted, together
with the numerical results and the kaon contribution.
\end{description}

\end{document}